\documentclass[useAMS,usenatbib]{mn2e}
\usepackage[pdftex]{graphicx}

\title[Outburst activity of symbiotic system AG~Dra]{Outburst activity of symbiotic system AG~Dra}
\author[L. Hric, R. G\'{a}lis, L. Leedj\"{a}rv et al.]
{L. Hric$^{1}$\thanks{E-mail: hric@ta3.sk}, R. G\'{a}lis$^{2}$, L. Leedj\"{a}rv$^{3}$, M. Burmeister$^{3}$ and E. Kundra$^{1}$\\
\\
$^{1}$Astronomical Institute, Slovak Academy of Sciences, Tatransk\'a Lomnica 059 60, The Slovak Republic \\
$^{2}$Faculty of Science, P. J. \v{S}af\'{a}rik University, Park Angelinum 9, Ko\v{s}ice 040 01, The Slovak Republic\\
$^{3}$Tartu Observatory, 61602 T\~{o}ravere, Estonia}
\begin{document}

\date{Accepted ??. Received ??; in original form ??}

\pagerange{\pageref{firstpage}--\pageref{lastpage}}
\pubyear{2014}

\maketitle

\label{firstpage}

\begin{abstract}
AG~Dra is a well known bright symbiotic binary with a white
dwarf and a pulsating red giant. The long-term photometry
monitoring and a new behaviour of the system are presented. The
detailed period analysis of photometry as well as spectroscopy
was carried out. In the system of AG~Dra, two periods of
variability are detected. The longer one around 550 days is
related to the orbital motion, and the shorter one around 355
days is interpreted as pulsations of the red giant in our older
paper. In addition the active stages change distinctively, but
the outbursts are repeated with the periods from 359 to 375
days.
\end{abstract}

\begin{keywords}
stars: binaries: symbiotic -- stars: individual: AG~Dra --
stars: oscillations
\end{keywords}

\section{Introduction}
AG~Dra is a classical symbiotic binary, type S. It is one of
the best studied symbiotic systems, thanks to its relatively
high brightness (about 9.8 mag in quiescence and about 8.3 mag
in the largest outbursts in $V$), high Galactic latitude ($b =
41^{\circ}$) favourable for observations and low extinction
($E_{(B-V)}=0.05$, \citet{mikolajewska_1995}).

The cool component of AG~Dra is a relatively early spectral
type with the classification in the range from K0 to K4, and of
low metallicity, [Fe/H] = $-1.3$ (Smith et al. 1996). Mass of
the giant was estimated on 1.5 $M_{\sun}$ by
\citet{kenyon_1987}. Its luminosity could be higher than that
of standard class III \citep{huang_1994, mikolajewska_1995}.
Smith et al. (1996) found the overabundance of elements heavier
than Fe (mainly Ba and Sr) and thus classified the giant as a
barium star. Such stars are on average more luminous than
standard giants of the same spectral class and they could be
more capable of invoking symbiotic activity in the binary
system due to their higher mass loss rate. The radius of the
giant was estimated to be $\sim 35 \,R_{\sun}$ by
\citet{zamanov_2007} and \citet{garcia_1986} found an orbital
separation of $400 \, R_{\sun}$. If we estimate the Roche lobe
radius on the basis of the mentioned values, we can conclude
that the giant probably does not fill its Roche lobe.

The hot component of AG~Dra is considered to be a white dwarf
sustaining a high luminosity ($\sim 10^3 \, L_{\sun}$) and
temperature ($\sim 10^5 \, \rm K$) due to the thermonuclear burning of
accreted matter on its surface \citep{mikolajewska_1995}. The
accretion most likely takes place from the stellar wind of the
cool giant. Both components are in a circumbinary nebula,
partially ionized by the hot component.

AG~Dra undergoes characteristic symbiotic activity with
alternating quiescent and active stages. Active ones consist of
several outbursts repeating at about a one-year interval. The
amplitudes of the outbursts decrease toward the longer
wavelengths, from $\sim 1$ mag in $V$ to $\sim 3$ mag in $U$.
Periodical outbursts and their relation to the orbital motion
of the binary system have been a matter of long-term debate.
While there is general agreement that the orbital period of
AG~Dra is about 550 days \citep{meinunger_1979, galis_1999,
fekel_2000}, there are variations on the shorter time scales
(350--380 days) presented by \citet{bastian_1998},
\citet{friedjung_1998, friedjung_2003} and some others.
Understanding the nature and mechanism of this variability is
crucial in order to explain the outburst activity of AG~Dra and
other classical symbiotic stars.

\citet{gonzalez_1999} showed on the basis of the analysis of
all \textit{IUE} observations that there are two types of
outbursts: the cool and the hot one. During the cool outburst
(e.g. 1981 - 83 and 1994 - 96), the hot compact object has a
temperature $\approx 90 \, 000 \,\rm K$ thus lower than that in
the quiescent stage. The hot outbursts (e.g. 1985 - 1986) are
characterized by temperatures above $\approx 130 \, 000 \,\rm
K$ and considerably lower optical and UV brightness.

\citet{skopal_2009} studied the supersoft X-ray vs. optical/UV
flux anticorrelation of AG~Dra. They concluded that such
behavior is caused by the variable wind from the hot component
of the system. \citet{shore_2010} found that variations in the
Raman features ratio, during the outburst, are consistent with
the disappearance of the O VI far UV resonance wind lines from
the white dwarf. The observations support the suggestion that
the soft X-ray and UV variations are due to the expansion of
the envelope and decreasing the effective temperature of the
gainer star. The outburst activity (2006 - 2008) of AG~Dra was
studied by \citet{munari_2009} using of the photometric and
spectroscopic observational material. The first outburst of
this activity stage was one of the cool type and was followed
by the fainter maximum of the hot type after 375 days.
\citet{contini_2011} mentioned that the collision of the wind
from the white dwarf with the dusty shells, ejected from the
red giant, leads to the fluctuations in $U$ band during the
major outburst. The long-term spectroscopic study of AG~Dra was
presented by \citet{shore_2012}. They discussed the effects of
the environment and orbital modulation in this system. Recently
\citet{leedjarv_2012} demonstrated spectroscopic behaviour of
AG~Dra before, during and after the last outburst.

Recently \citet{formiggini_2012} accomplished complex period
analysis of the historical optical light curve of AG~Dra,
covering the last 120 years. They found that the period of the
outbursts was 373.5 days and a cyclical behaviour with a
quasi-period of 5300 days. The last one is the time interval
between the start of the active stages induced by the
solar-like cycles of magnetic activity. The combined effect of
the 5300 and 373.5 days cycles augments the postulated cool
giant pulsations of the red giant with the period of 350 days
\citep{formiggini_2012}. However, the proposed rotational
period of the giant of 1160 days in their model, and in
particular, its retrograde rotation seems to be somewhat
artificial and not well justified physically.

In the present paper, we re-analyze the $UBVR$ light curves of
AG~Dra and also apply time series analysis to the spectroscopic
data. We use all the available photometry, starting from the
compilation of photographic observations by
\citet{robinson_1969} and proceeding with a large amount of
photoelectric observations (see hereafter). Variability of the
emission lines provides an additional information of the
physical state of the AG~Dra system.

\section[]{Photometric and spectroscopic data}
The new photoelectric and CCD observational
material\footnote{The photometric data are available upon
request from the authors.} was obtained at observatories at
Skalnat\'{e} Pleso (SP), Star\'{a} Lesn\'{a} (SL-G1 and SL-G2)
and Vala\v{s}sk\'{e} Mezi\v{r}i\v{c}\'{i} (VM). At the
observatories SP and SL-G2 identical Cassegrain telescopes with
a diameter of 0.6 m were utilised. One-channel photoelectric
photometers with digital converters were used, as well as
standard UBVR Johnson's filters. CCD photometry was performed
at SL-G1 using the 0.5-m telescope. The SBIG ST10 MXE CCD
camera with a chip of $2184 \times 1472$ pixels and the $ \rm
UBV(RI)_{\rm C}$ Johnson-Cousins filter set were mounted at the
Newtonian focus.
\begin{table}
\begin{center}
\scriptsize
\begin{tabular}{cccc}
\hline
Observatory & $ N_{\rm nights} $ & Period & Period [MJD]\\
\hline
SP + SL-G2   &  & $ 18. 5. 1999 - 16. 3. 2012 $ & $51\,317 - 56\,003$ \\
SL-G1        &  & $ 31. 7. 1994 - 27. 9. 2009 $ & $49\,565 - 55\,102$ \\
VM           &  & $ 24. 4. 1997 - 24. 4. 2005 $ & $50\,563 - 53\,485$ \\
\hline
\end{tabular}
\caption[]{List of observational nights on particular
observatories: SP - Skalnat\'{e} Pleso, SL-G1 - Star\'{a}
Lesn\'{a}, pavilion G1, SL-G2 - Star\'{a} Lesn\'{a}, pavilion
G2 and VM - Vala\v{s}sk\'{e} Mezi\v{r}i\v{c}\'{i}. }
\label{observations} \normalsize
\end{center}
\end{table}

Schmidt-Cassegrain (280/1765 mm) equipped with CCD ST-7 and set
of VR filters was used at VM observatory. From $\rm JD \,2 \,
450 \, 563$ to $\rm JD \,2 \, 451 \, 179$ photographic films as a
detector were used. The number of observational nights,
observational intervals in the date as well as in Julian days
for particular observatories are listed in Table
\ref{observations}. In addition, we used the same data that had
already been analyzed and discussed in our previous paper
\citep{galis_1999} and published photoelectric UBV photometry
by \citet{belyakina_1965, belyakina_1969}, \citet{skopal_2002},
\citet{leedjarv_2004}, \citet{skopal_2004},
\citet{munari_2009}, \citet{skopal_2012} as well as
photographic measurements by \citet{robinson_1969} and
\citet{luthardt_1983}.

To support the model of cool giant pulsations we obtained the
photometric observations in instrumental $\Delta \rm R_{\rm i}$
magnitude. These magnitudes were not transformed to the
international system due to the lack of the comparison star $R$
magnitudes. These observations were secured in the period from
JD 2 449 557 (July 23, 1994) to JD 2 455 102 (September 27,
2009) at the SP and SL-G1 observatories and part of them was
published by \citet{skopal_chochol_1994}, \citet{skopal_1995},
\citet{hric_1996}, \citet{skopal_1998}, \citet{skopal_2002,
skopal_2004, skopal_2012}.

Intermediate dispersion spectroscopy of AG~Dra was carried out
at the Tartu Observatory in Estonia. The data used in this
paper covered the interval from JD 2 450 702 (September 11,
1997) to JD 2 455 651 (March 31, 2011). All together 211
spectra of AG~Dra were used for the period analysis. The
spectroscopic material was taken by the 1.5 m telescope
equipped with the Cassegrain grating spectrograph. The
equipment and method of observations were described in
\citet{leedjarv_2004} and \citet{leedjarv_2012}. The majority
of the spectra were recorded in two spectral regions which we
call red and blue. Red spectra were taken with a dispersion of
0.47 \AA /pix at $\rm H_{\rm \alpha}$~and they include emission
lines of $\rm H_{\rm \alpha}$, He I 6678 \AA~and the Raman
scattered O VI line at 6825 \AA. The blue spectra included at
least, emission lines of $\rm H_{\beta}$, He~II~4686 \AA \, and
He~I~4713 \AA. The dispersion of the blue spectra was about
0.57 \AA /pix at $\rm H_{\rm \beta}$. The spectra were reduced
using the software package MIDAS provided by \textit{ESO}. The
wavelength-calibrated spectra were normalized to the continuum
and positions, peak intensities and equivalent widths (EW) of
the emission lines were measured\footnote{The spectroscopic
data are available upon request from the authors.}.
\begin{figure}
\begin{center}
\includegraphics[width=8.4cm]{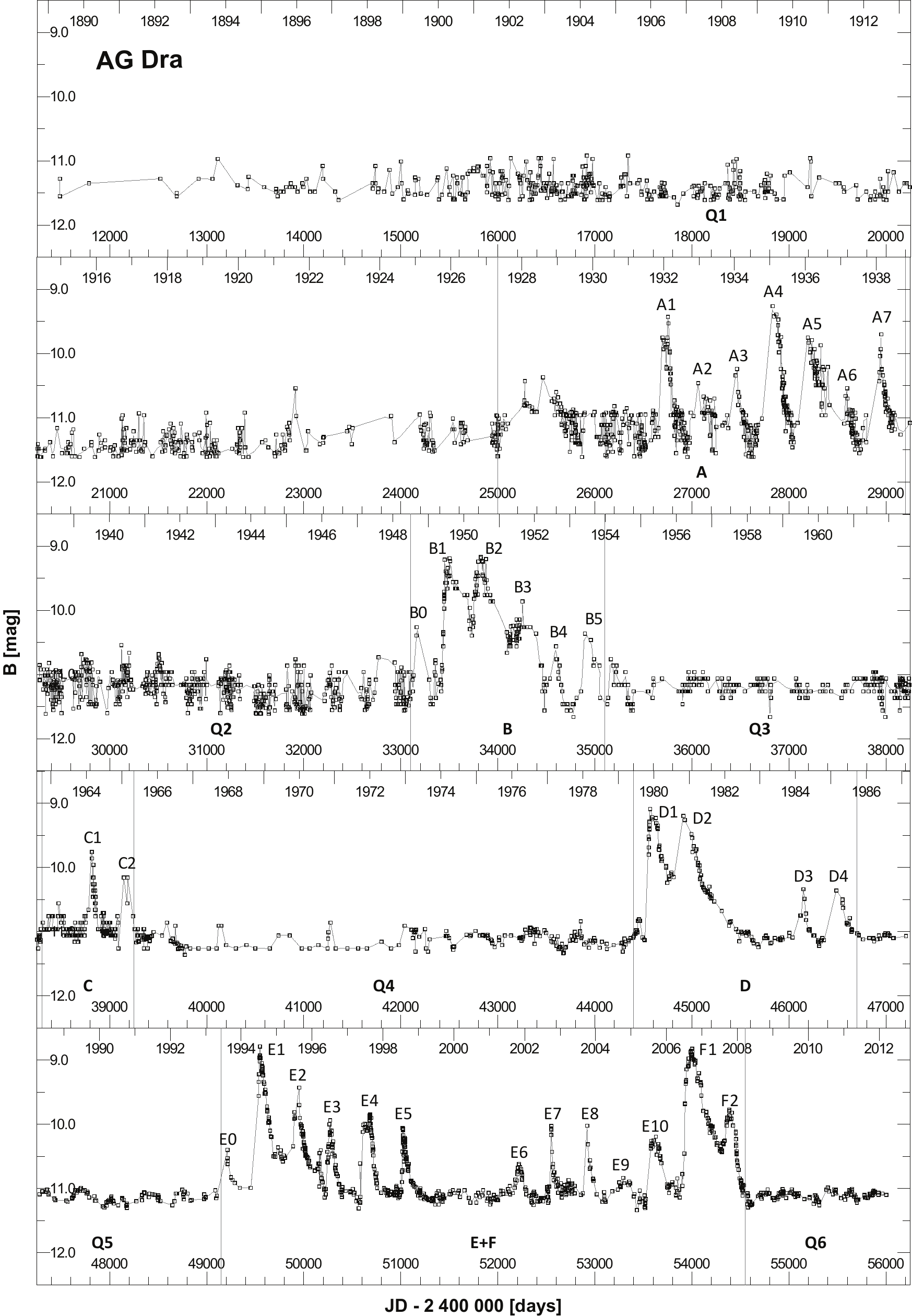}
\caption{The historical LC of AG~Dra from period 1889 - 2012
constructed on the basis of photographic and $B$ photoelectric
observations. The used data sources are quoted in the text. The
LC is devided to active (A - F) and quiescence stages (Q1 - Q6)
by vertical lines. The thin curve shows a spline fit to the data points.}
\label{ag_b}
\end{center}
\end{figure}
\section[]{Morphology of the light curves and period analysis}
In this section, there are described the results of re-analysis
of the historical light curve from the photographic data up to
1966. We have also presented the results of the analysis of
UBVR photometry from 1974 to the present day. The results of
principal component analysis of UBV photometry are also
discussed.

The period analysis of the observational data was performed
using an advanced implementation of the Date Compensated
Discrete Fourier Transform. We used a Fisher Randomization Test
(Monte Carlo Permutation Procedure) to calculate two
complimentary False Alarm Probabilities (FAP) for determining
the significance of a given period $P$: FAP 1 represents the
proportion of permutations containing a period with a
peak/valley higher (respectively lower) than the peak/valley of
$P$ at any frequency. It is the probability that there is no
period in the period window (power spectrum) with value $P$.
FAP 2 represents the proportion of permutations containing a
period with a peak/valley higher (respectively lower) than the
peak/valley of $P$ at exactly the frequency of $P$. It is the
probability that the observation data contain a period that is
different from $P$. This test executes the selected period
analysis calculation repeatedly (at least 100 times), each time
shuffling the magnitude values of the observations into the
form of a new randomized observation set, but keeping the
observation times fixed \citep{press_1992}. FAPs with value
below 0.01 (1\%) mostly indicate the significant periods. FAPs
above 0.20 (20\%) mostly relates to an artifacts in data,
instead of a true period.

In addition, we created a spectral window to confirm that the
periods found in the previous steps are not artifacts of the
observing rate. We performed the detail period analysis for
some complicated cases (e.g. during active stages), in which
the response of examined period $P$ was removed from the data
and the power spectra of such residuals were studied with a
focus on persistent presence of aliases and harmonics of this
period. The minimum period error (or period uncertainty) of the
period $P$ was determined by calculating a 1-sigma confidence
interval on $P$, using a method described by
\citet{schwarzenberg_1991}.

\subsection{The light curve between the years 1890 and 1966}
The first historical photometric observations were dated to the
end of 19th century (Fig. \ref{ag_b}). During the period (1890
- 1996) the AG~Dra system underwent 3 phases of activity: the
first one between the years 1932 and 1939, the second one
between 1949 and 1955 and the third one between 1963 and 1966.
In total, we recognised  15 outbursts in this period: seven
during the first active phase, six during the second one and
two outbursts during the last one.

\begin{figure}
\begin{center}
\includegraphics[width=0.45\textwidth]{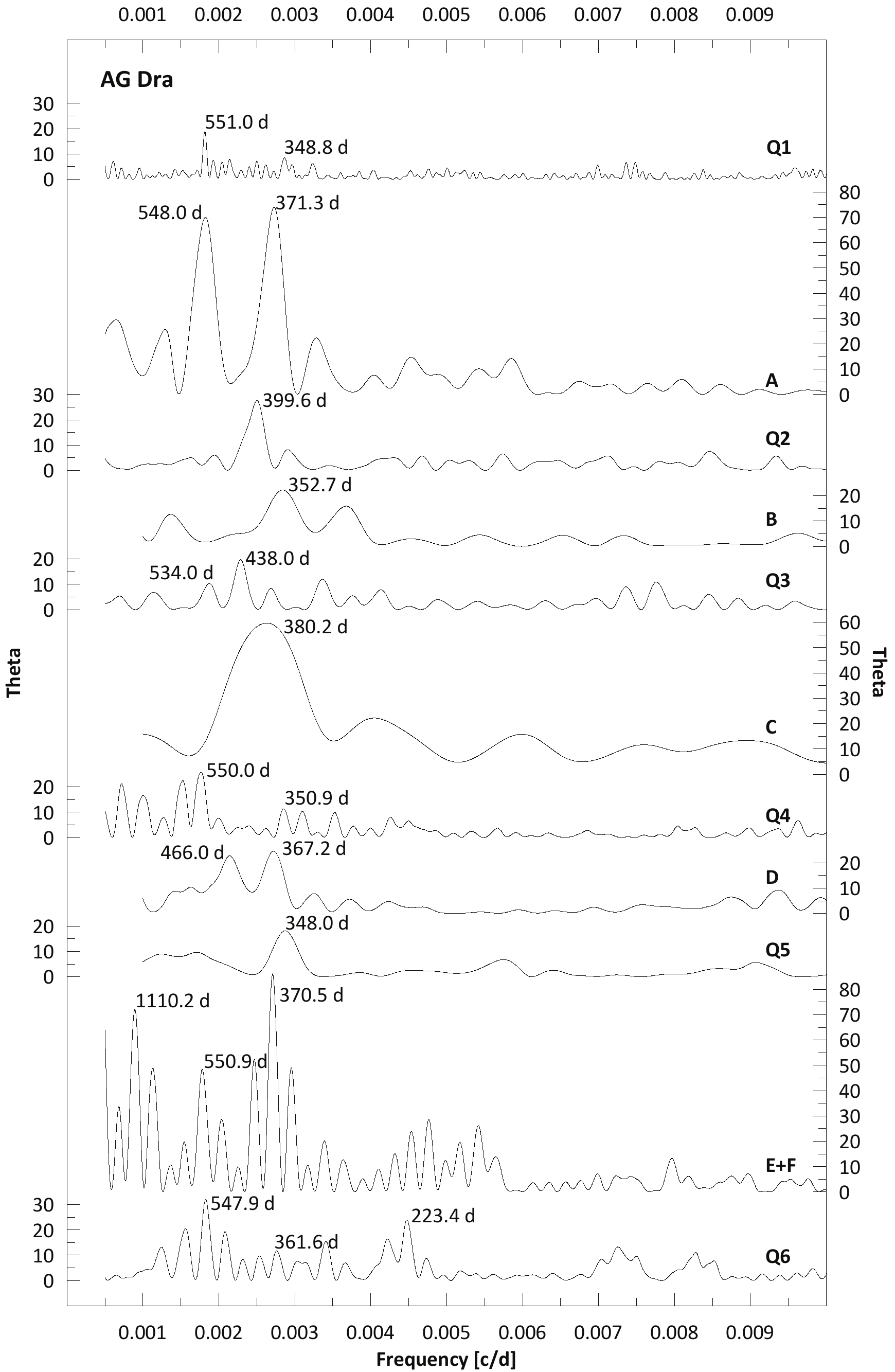}
\caption{Power spectra of AG Dra taken from historical
photographic as well as photoelectric and CCD data in $B$
filter for particular stages of quiescence (Q1 - Q6) and
activity (A - F). Significant periods are marked with their
values.}
\label{ag_perB}
\end{center}
\end{figure}

We present the results of the re-analysis of this data with
contemporary method described in Section 3. We divided the
light curve into quiescent and active stages (marked by
vertical lines in Fig. \ref{ag_b}) and carried out a period
analysis independently for each stage. We present the results
of this analysis for the given stages in Table
\ref{periods_hist} as well as in Fig. \ref{ag_perB}. We can
conclude that the historical light curve shows both known
periods: $\approx 550 \,\rm days$ (orbital period) and $\approx
350 \,\rm days$ (the period of postulated pulsation of the red
giant, \citet{galis_1999}). Besides these periods the analysis
gave us the period 370 - 380 days which is present in the
active stages A and C. This period is related to the recurrence
of the individual outbursts occurred in these active stages.
The periods around 400 days (Q2) and 440 days (Q3) are probably
not real and are caused by the low amplitude of light
variations in comparison with the scatter of these photographic
data.

\begin{table}
\begin{center}
\scriptsize \caption{The results of the period analysis of
historical light curve of AG~Dra constructed on the basis of
photographic and $B$ photoelectric observations. $T_{\rm
start}$ marks the beginning and $T_{\rm end}$ the end of the
given stage. During stages Q1, Q2 and Q3 was the system
quiescent whereas stages A, B and C are active.}
\begin{tabular}{cccc}
\hline
Phase & $T_{\rm start}$ & $T_{\rm end}$ &      Significant periods         \\
      &    [MJD]        &     [MJD]     &           [days]                 \\
\hline
Q1    & $ 11\,500$      & $ 25\,000$    & $551.0 \pm 3.5; 348.8 \pm 2.1$   \\
A     & $ 25\,500$      & $ 29\,200$    & $371.3 \pm 4.6; 548.0 \pm 8.1$   \\
Q2    & $ 29\,200$      & $ 33\,100$    & $399.6 \pm 5.6$                  \\
B     & $ 33\,100$      & $ 35\,100$    & $352.7 \pm 5.9$                  \\
Q3    & $ 35\,100$      & $ 38\,300$    & $438.0 \pm 10.7; 534.0 \pm 23.9$ \\
C     & $ 38\,300$      & $ 39\,250$    & $380.2 \pm 10.2$                 \\
Q4    & $ 39\,250$      & $ 44\,400$    & $550.0 \pm 10.3; 350.9 \pm 4.8$  \\
\hline
\end{tabular}
\label{periods_hist}
\end{center}
\end{table}

Between the years 1889 and 1927 (stage Q1) AG~Dra was in
quiescence with a mean photographic brightness 11.02 mag. Small
variations of the LC revealed the presence of both known
periods. After this, at least 38 years of quiescence, the
system went into an active stage (assigned by A in Fig.
\ref{ag_b}). Until the year 1938 we can recognize seven major
outbursts occurring approximately every 300 - 400 days. The
individual outbursts have a rather steep rise to the maximum
and a fast decline. The outburst in the year 1932 (A1) had two
maxima with a second (more prominent) maximum following the
first one after $\approx 50 \,\rm days$. The system reached the
maximal brightness of 8.9 mag during the fourth outburst (A4).
For the next almost 11 years a quiescent stage followed with
only small semi-regular variations. The period analysis gave us
a period with a value $\approx 400 \,\rm days$, which is more
probably related to the data distribution in this part of the
LC (see stage Q2 in Fig. \ref{ag_b}) as the real brightness
variability of AG~Dra system. This distribution might have
drowned the expected orbital or probable pulsation period.
An epoch of intense activity was observed between the years
1949 and 1955 (stage B). The overall morphology of the LC
during this active stage is rather different from that of stage
A. The peaks have an analogously steep rise to the maximum with
the highest achieved brightness 8.8 mag. The decrease is
considerably slower and unlike stage A, the system spends most
of this period above 10.5 mag. The outbursts follow one after
another $\approx 350 \,\rm days$. The second and third
outbursts (B2 and B3) might have again secondary maxima
approximately 50 days after the primary ones.

Stage Q3 (1955 - 1963) is characterised by the semi-regular
featureless brightness variations with period $\approx 440
\,\rm days$. This behaviour probably influenced the value of
the orbital period (534 days) obtained by our analysis . The
system became active after the year 1963, which resulted in the
shortest (1963 - 1966) and the least active (only 2 - 3
outbursts) stage during whole observed LC of AG~Dra. The period
analysis of this active stage C revealed the presence of 380
days period.
\begin{figure*}
\begin{center}
\includegraphics[width=\textwidth]{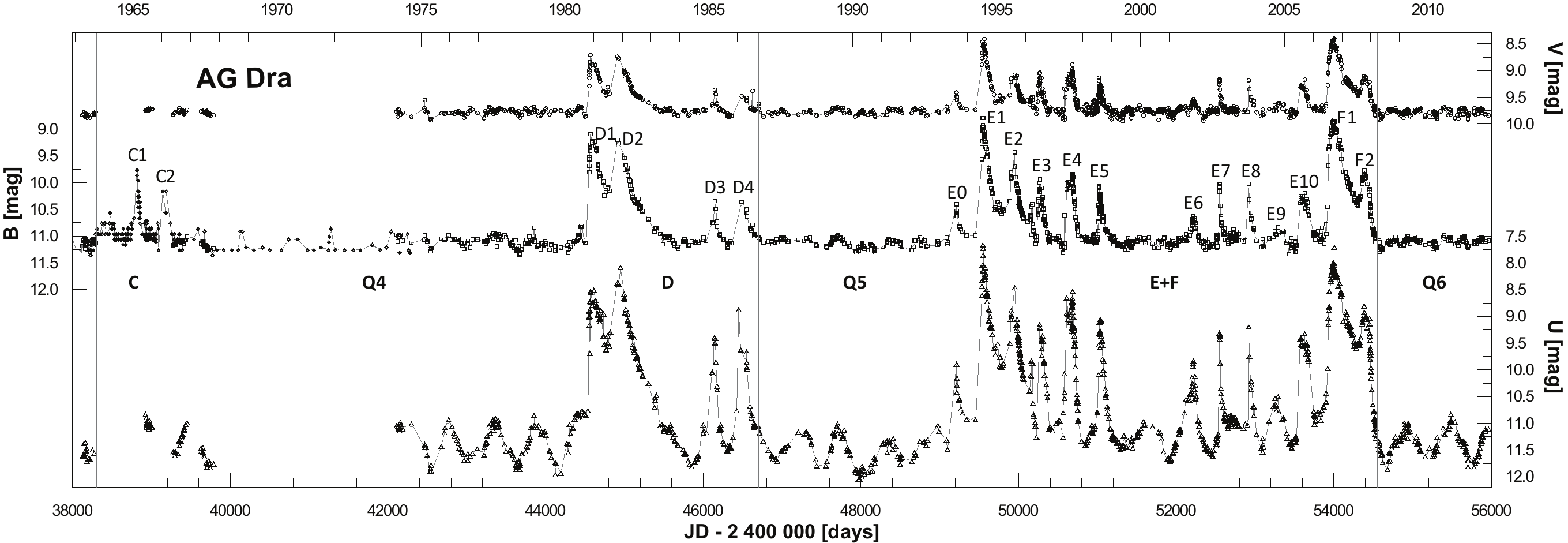}
\caption{UBV LCs from the period 1963 - 2012 with marked active
stages (C, D, E and F) and quiescent ones (Q4, Q5 and Q6).
Particular outburst are assigned as C1 - C2, D1 - D5, E0 - E10
and F1, F2.  The thin curves show a spline fits to the data
points.} \label{UBV}
\end{center}
\end{figure*}
\subsection[]{The light curve after 1966}
For complete coverage of the 124 year history of AG~Dra we are
missing an 8 years part on the LC between $\rm JD \, 2 \, 439
\, 355$ (August 17, 1966) and $\rm JD \, 2 \, 442 \, 109$
(March 2, 1974). In the archive of the Sonnenberg Observatory
we have found two papers \citep{splittgerber_1974,
luthardt_1983} where the LC behaviour during this period is
described, but the data was not available. We extracted the
photometric data from the LC depicted in Fig. 2 in the paper by
\citet{luthardt_1983} to cover the missing part of the
historical LC. It was obvious from the completed LC (Fig.
\ref{ag_b}) that after the outburst C the system fell into the
long-term quiescent stage (assigned as Q4). The activity of the
system was very low during this period and the average
photographic brightness was 10.9 mag. The period analysis of
this part of the LC shows unambiguously the presence of 550
days orbital as well as 350 days postulated pulsation period.

\begin{table*}
\begin{center}
\scriptsize \caption{The results of period analysis of
particular stages between 1963 - 2012. LCs for each filter were
analyzed separately. $T_{\rm start}$ is beginning and $T_{\rm
end}$ the end of the given stage. The periods are in order
according to their significance.}
\begin{tabular}{cccccc}
\hline
Phase    & $T_{\rm start}$ & $T_{\rm end}$ & \multicolumn{3}{c}{Significant periods [days]} \\
         &     [MJD]      &    [MJD]     & $U$ band      & $B$ band                                          & $V$ band \\
\hline
Q4       & $ 39\,250$     & $ 44\,400$   & $551.0 \pm 2.4$ & $550.0 \pm 10.3; 350.9 \pm 4.8$                     & $349.6 \pm 13.9; 550.0 \pm 49.7$ \\
D        & $ 44\,400$     & $ 46\,700$   & $371.9 \pm 5.5$ & $367.2 \pm 8.1; 466.0 \pm 15.2$                     & $372.5 \pm 6.1$ \\
Q5       & $ 46\,700$     & $ 49\,150$   & $553.6 \pm 4.0$ & $348.0 \pm 6.7$                                     & $350.1 \pm 7.3$  \\
E + F    & $ 49\,150$     & $ 54\,550$   & $371.2 \pm 1.8$ & $370.5 \pm 1.9; 1\,110.2\pm18.5^{a}; 550.9 \pm 9.8$ & $370.5 \pm 1.8$  \\
Q6       & $ 54\,550$     & continue     & $549.3 \pm 2.7$ & $547.9 \pm 6.4; 223.4\pm5.3^{b}; 361.6 \pm 5.3$     & $357.3 \pm 19.8$ \\
\hline
\multicolumn{6}{l}{Notes:} \\
\multicolumn{6}{l}{a - The period of $1\,110.2\pm18.5$ days is probably
only double of $550.9\pm9.8$ days one.} \\
\multicolumn{6}{l}{b - The period of $223.4\pm5.3$ days is 1-year alias
of $547.9\pm6.4$ days one.} \\
\multicolumn{6}{l}{The global morphology of active stages D and E+F is
possible very well describe by sinusoidal variations} \\
\multicolumn{6}{l}{with periods around $1\,110$ days (or double $2\,220$ days)
and $2\,500$ days (or double $5\,000$ days).} \\
\end{tabular}
\label{periods_UBV}
\end{center}
\end{table*}

The quiescent stage Q4 is partly covered by photoelectric
observations, too. The first photoelectric UBV photometry of
AG~Dra was obtained by Belyakina from 1962 to 1967
\citep{belyakina_1965, belyakina_1969}. Since 1974 the system
of AG~Dra has been observed systematically mainly
photoelectrically in {UBV}. The merit of good coverage of the
LC has been our campaign \citep{hric_skopal_1989}, too. The
benefits of this data were the improvement of the orbital
period ($\approx 550\,\rm days$) and the discovery of the
shorter period ($\approx 350\,\rm days$), which was interpreted
as the pulsation of the cool component \citep{galis_1999}. The
photoelectric LCs of AG~Dra are depicted in particular filters
in Fig. \ref{UBV}. It is obvious from the LCs in $U$, $B$ and
$V$ bands, that the amplitudes of variations decrease towards
longer wavelengths. During the activity of the system the LCs
are in strong correlation in all bands. The variations in
quiescence are similar in $B$ and $V$, but differ in $U$ bands.

In 1980 the system entered the active stage D, where outbursts
were characterized by fast increase and slower decrease as in
stage B. The individual outbursts are clearly recognised in
Fig. \ref{UBV}. It is worth noting that the active stage D was
interrupted by the interval (1983 - 1985) when the system
manifested the behaviour typical for quiescent stages. AG~Dra
was in quiescent stage (Q5) for the next 7 years. In 1993 new
activity (E) started with the small outburst followed by the
next 5 prominent outbursts consecutively in $\approx 350 -
390\,\rm days$ intervals. Activity of the system was again
interrupted by a short quiescent-like interval in 1999 - 2001
(similar to the behaviour happened during the active stage D)
and continued by the next 5 outbursts which finished at the
beginning of 2006. Such long-term activity has never been
observed during more than one century of photometric history of
AG~Dra. This activity was immediately (without quiescence
phase) followed by the large outburst (F1) in 2006 when the
brightness in $U$ reached 8 mag. This new phase of activity (F)
had two (up to $\rm JD \, 2\, 455\, 000$) outbursts followed by
a fast drop to the deep brightness minimum (11.8 mag in $U$
band). Since 2008 the symbiotic system AG~Dra has been in
quiescent stage (Q6).

From the light curve in Fig. \ref{UBV}, we can distinguish two
kinds of outbursts. There are main double-peaked outbursts (B1
and B2, D1 and D2, E1 and E2, F1 and F2) with very rapid
increase of brightness, small drop of magnitude between peaks
and slow decrease to the quiescence level. The identification
of double-peaked outburst is ambiguous during the active stage
A and such a feature is totally missing in the weak active
stage C. This type of outbursts defines the beginning of each
new cycle of activity of the AG~Dra with time interval 12 - 16
years. Moreover, there are outbursts with the sharper shape and
lower amplitude which are observed after main outburst and
sometimes during the whole cycle of activity (B3 - B5, C1 - C2,
D3 - D4, E3 - E10).

Our statistical analysis of the photoelectric and CCD
observations shows that the light curves in $U$, $B$ and $V$
filters were very well correlated (correlation coefficients
$\approx 0.9$) during the active stages (D, E and F in Fig.
\ref{UBV}). During the quiescent stages (Q4 - Q6 in Fig.
\ref{UBV}), the correlation coefficient of the LCs in band $U$
and $B$ as well as one of LCs in band $U$ and $V$ are less than
0.5. On the other hand the variations in $B$ and $V$ bands are
correlated quite well during the quiescence. This result showed
that the brightness variations during quiescent stages of
AG~Dra in the various bands were caused by different physical
mechanisms.

\begin{figure}
\begin{center}
\includegraphics[width=0.45\textwidth]{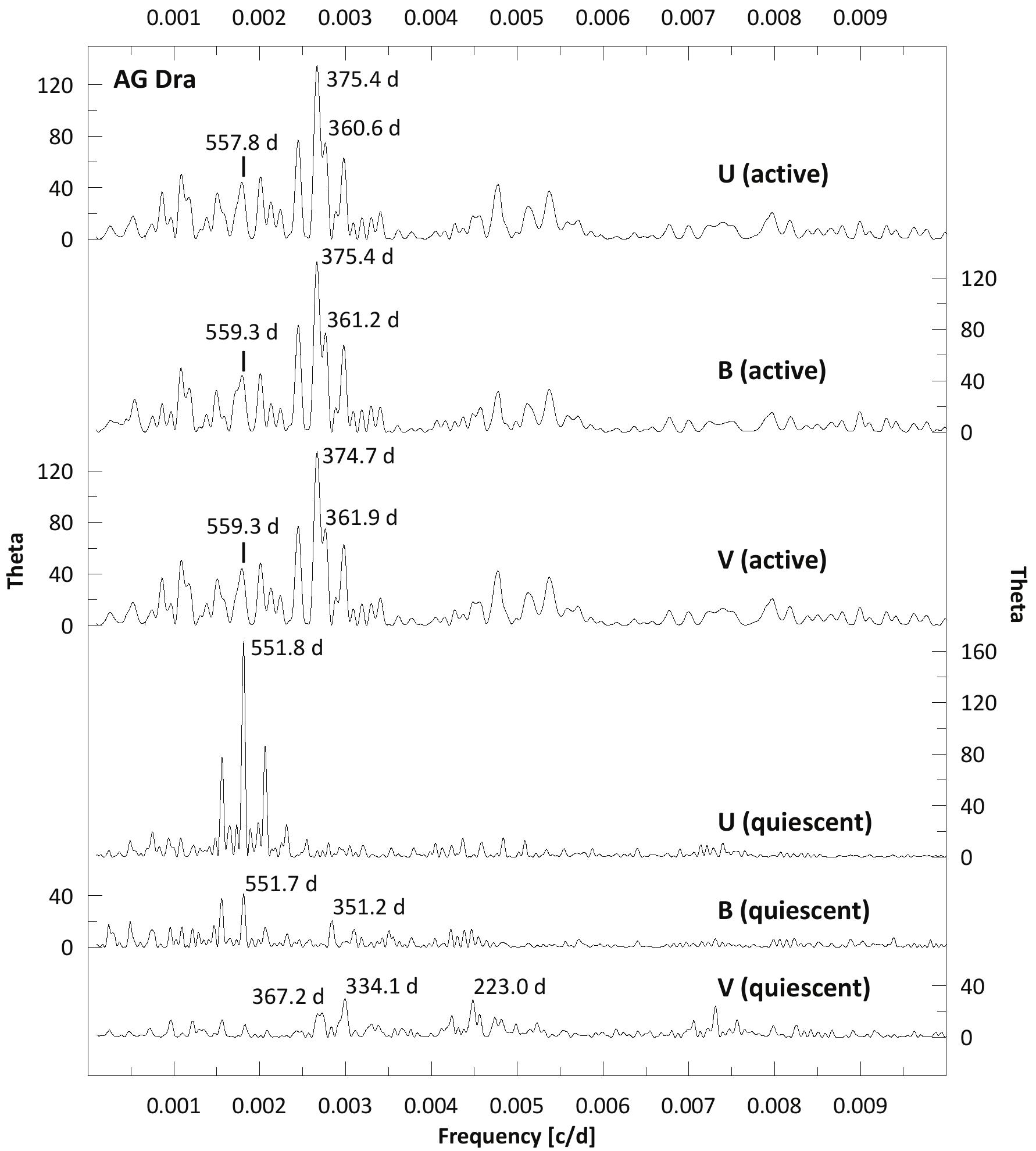}
\caption{Power spectra of AG Dra taken from photoelectric and CCD data in $U$, $B$ and $V$ filters for active (D, E+F)
and quiescent (Q4 - Q6) stages. Power spectra for active stages in $U$, $B$ and $V$ filters were obtained after
removing of long-term periods around 1500 and 5400 days, which are related to the global morphology
of these active stages.}
\label{AG_perUBV}
\end{center}
\end{figure}

We performed the period analysis of LCs in individual bands
($U$, $B$ and $V$) with the same methods used for period 1890 -
1966. The results of this analysis are presented in Table
\ref{periods_UBV}. The power spectra of AG Dra in $U$ , $B$ and
$V$ filters for active stages as well as for quiescence are
depicted in Fig. \ref{AG_perUBV}. As we have shown by our
period analysis of the quiescent stages, the light curve in $U$
band is clearly dominated by variations with the orbital period
$\approx 550\,\rm days$. In the bands $B$ and $V$, we found the
shorter period ($\approx 350\,\rm days$); although, its value
in each quiescent stage changed slightly. The period analysis
of the active stages revealed many significant periods. Our
detailed analysis showed that most of the periods were more
likely related to the complex morphology of the light curves
during the active stages than the real variability present in
this symbiotic system.

The significant period with a value of around 370 days is
related to the distribution of individual outbursts, which
dominate the light curve in the active stages. It should be
noted, however, that the value of this period varies with
wavelength and is different for the individual active stages.
Statistical analysis of the outburst distribution shows that
the median of the time interval between the individual
outbursts is 365 days, while the time intervals vary from 300
to 400 days without apparent long-term trend.

We can conclude, that the results of period analysis of UBV LCs
strongly confirm both known periods. The dominance of the
orbital period in $U$ filter is in agreement with the model by
\citet{galis_1999}.

Stronger influence of the second period in longer wavelength
suggests its relation with the behaviour of the red giant and
probably is caused by its pulsations as it was proposed by us
in the paper mentioned above.

\subsection[]{Differential $R_{\rm i}$ photometry of AG~Dra
during 1994 - 2009}
As we have shown in the previous section, two periods are
present in the LCs of AG~Dra. For the 350 day period
\citet{galis_1999} proposed the interpretation as pulsations of
the cool giant. To support this model we carried out the period
analysis of photometric observations in instrumental $\Delta
R_{\rm i}$ magnitude.

\begin{figure}
\begin{center}
\includegraphics[width=8.4cm]{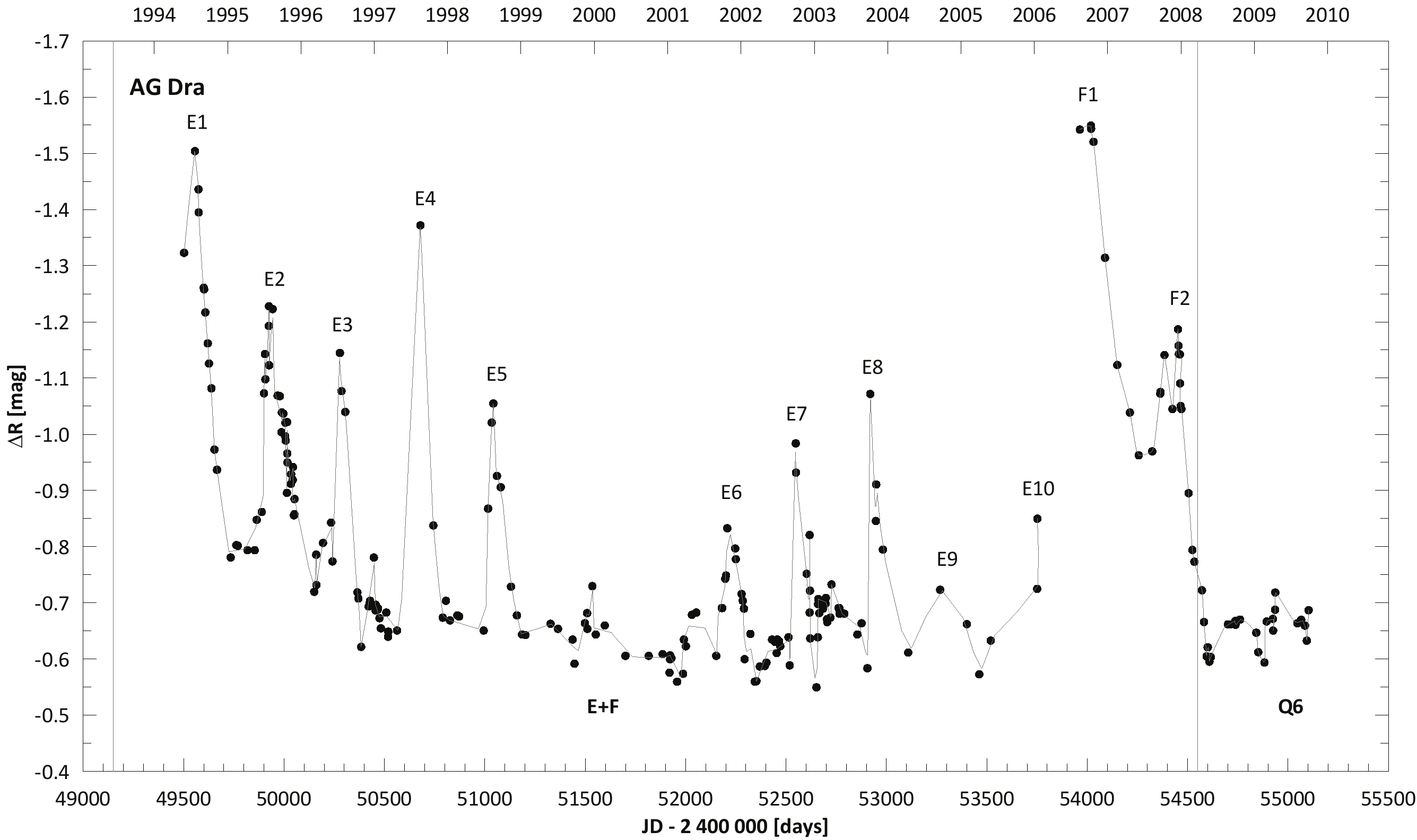}
\caption{Light curve of AG~Dra in instrumental $\Delta \rm R_{\rm i}$ mag.
The thin line shows a spline fit to the data points.}
\label{LC_R}
\end{center}
\end{figure}

We had at our disposal, 226 nights of observations covering
part of the active phase (E in Fig. \ref{UBV}). All
observations are depicted in Fig. \ref{LC_R}. Pronounced
outbursts as well as variations during quiescence are modulated
with the period $P_{\rm R} = (371.0 \pm 2.2)$ days. The
presence of this period at longer wavelength during quiescence
of AG~Dra probably suggests its relation to the cool component
in this system. The red giant pulsations proposed by
\citet{galis_1999} are the possible source of modulations with
this period.
\subsection[]{Results of principal components analysis of UBV photometry}
We accomplished the principal component analysis (PCA) of UBV
photometry from the period 1974 - 2006. We analysed the whole
light curve and then particular stages (Q4, D, Q5 and E)
separately. PCA is a type of linear orthogonal transformation
(also known as discrete Karhunen-Lo\`{e}ve transform or
Hotelling transform) to convert a set of observations of
possibly correlated variables into a set of values of linearly
uncorrelated variables, called principal components. This
transformation is defined in such a way that the first
principal component (PC1) has the largest possible variance,
and each succeeding component in turn has the highest variance
possible under the constraint that this component is orthogonal
to the preceding ones.

\begin{figure}
\begin{center}
\includegraphics[width=0.45\textwidth]{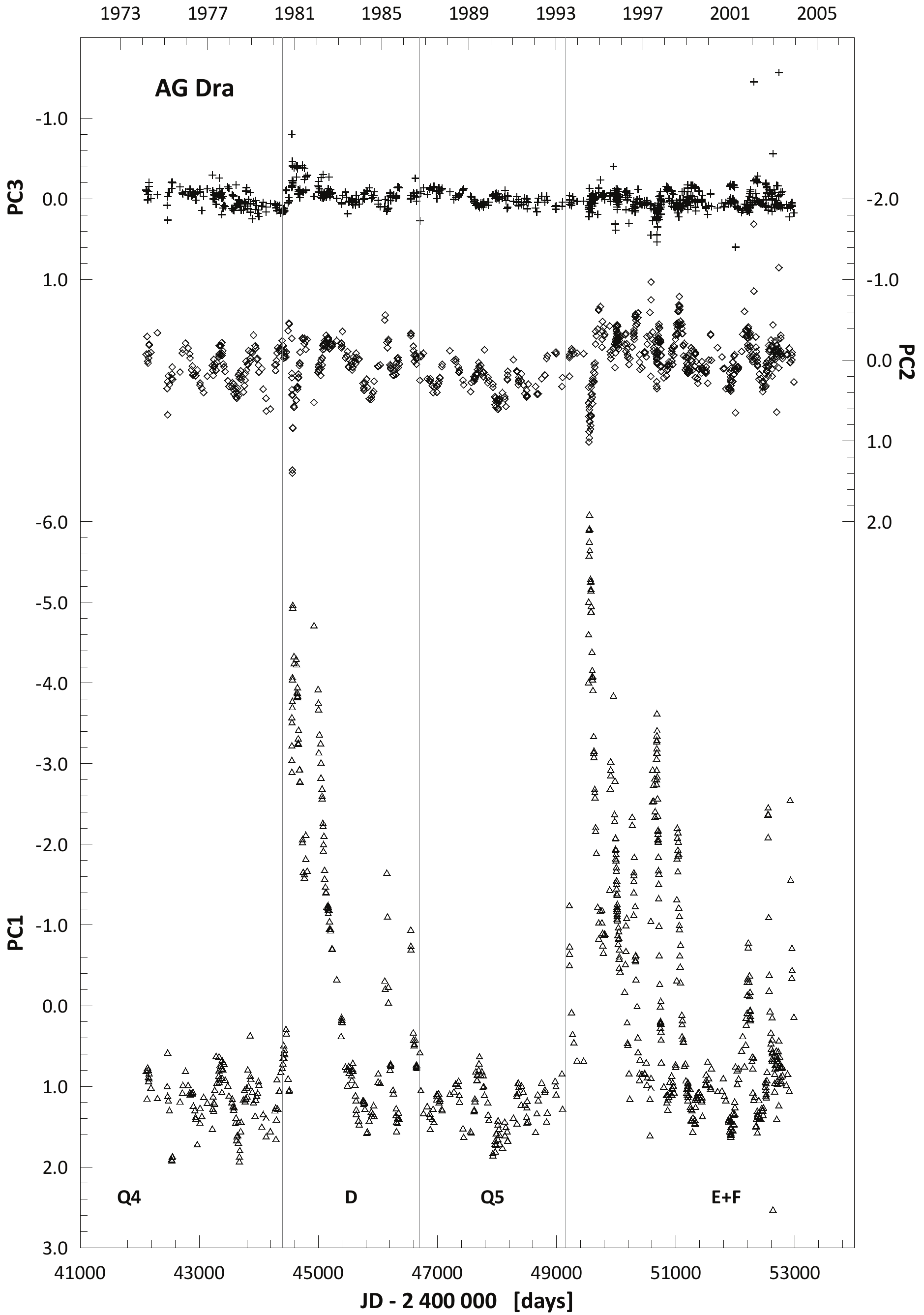}
\caption{Principal components PC1, PC2 and PC3 for AG~Dra UBV light curves.}
\label{PCA}
\end{center}
\end{figure}

For this analysis, only the observations with magnitudes in all
three filters were used.There were composed the input values of
three vectors $U \equiv (U_{\rm t1},U_{\rm t2},..., U_{\rm
tN})$, $B \equiv (B_{\rm t1},B_{\rm t2},... ,B_{\rm tN})$ and
$V \equiv (V_{\rm t1}, V_{\rm t2},..., V_{\rm tN})$, where
$U_{\rm ti}, B_{\rm ti} \,\,\rm and \,\, V_{\rm ti}$ were
magnitudes in particular filters in time $t_{\rm i}$ and $N$ is
number of observations. The output of the method is ternary of
vectors PC1, PC2 and PC3, which represents principal components
of input observations. They have $N$ dimensions as the input
vectors and therefore it is possible to assign time $t_{\rm i}$
and create the light curves of principal components (see Fig.
\ref{PCA}). The diagrams in this figure represent the light
curves of principal components computed for the period 1974 -
2006. If we compare the components with the original light
curves, we can see that the first component PC1 includes the
most of the significant variations in original data. The second
component PC2 contains the variation related to the orbital
motion. The last component PC3 does not contain significant
information about the binary system, only the noise caused by
the errors of the data.

The informational content in the particular components is
possible to quantify by variation analysis. The amplitudes of
variations in $U$, $B$ and $V$ bands decrease towards the
longer wavelengths. We scaled the original $U$, $B$ and $V$
light curves to unit variations to secure the same weights for
the LCs inputting into the analysis. It is evident from the
Table \ref{varPrincComp}, that output components are not
equivalent. The variation of PC1 composes as far as 96.3 \% of
the variation of input data and so represents 96.3 \%  of
entire information content. That is caused by the mechanism of
PCA which selects the principal component PC1 so that its
correlation with the input signal is the highest. The whole
variation of principal components is equal to the whole
variation of input data. It means that the principal components
contain whole information about input light curves.

In the next step the light curves were divided into the
particular stages (Q4, D, Q5, and E) and the variation analysis
was accomplished for each stage separately. This analysis gave
similar results as in the case of the whole LCs.

Apart from variation analysis of particular components, it is
interesting to study their correlations as well as the
correlations to the original light curves. We analysed
correlations among all combinations of light curves ($U$, $B$,
$V$) and principal components (PC1, PC2, PC3). Very interesting
are the correlations of the light curves in particular filters.
During the activity these correlation coefficients  are very
high $(\geq 0.97)$. The behaviour of $U$, $B$ and $V$ light
curves are equivalent, because outbursts are dominating
features in all bands. In this manner, we detected high
correlation with PC1, too. The behaviour of quiescent stages
are different. The light curves in $U$ and $B$ bands have a
correlation coefficient $\approx 0.7$, but in $U$ and $V$ bands
have a correlation coefficient only $\leq 0.5$, because the
orbital period is dominating in $U$ band while the probable
pulsation period is more significant in $V$ band.

The brightness variations related to the orbital motion as well
as the probable red giant pulsation should be present in all
three bands, but with different amplitudes. The method of PCA
transforms the variations into the first principal component.
Moreover, the advantage of PC1 period analysis in comparison
with analysis in particular colours is such, that the first
principal components contain the most of the informations, the
noise is suppressed (is isolated in the last component) and the
detected periods have higher statistical significance.
Therefore we accomplished the period analysis of PC1. The
result of this analysis for particular stages are listed in
Table \ref{varPrincCompPC1}. The principal component PC1
contains two significant periods around 550 and 350 days, even
the individual values of detected periods vary from 343 to 368
days and from 526 to 576 days for postulated pulsation and
orbital periods, respectively. The presence of 368 days period
is possible to explain by the time distribution of individual
outbursts during active stages. The large scatter in the values
of orbital period (larger than the typical error of period
determination) as well as the presence of 485 days period is
not clear.

\begin{table}
\centering \caption{Variation analysis of principal components
of whole period (1974 - 2006)}
\begin{tabular}{lll}
\hline
Component & Variation & Cumulative sum \\
\hline
PC1 & 2.889 & 2.889 \\
PC2 & 0.089 & 2.979 \\
PC3 & 0.021 & 3.000 \\
\hline
\multicolumn{2}{l}{The original light curves:} & 3.000 \\
\hline
\end{tabular}
\label{varPrincComp}
\end{table}

Fig. \ref{PCA} suggests that the principal component PC2
contains only the orbital variations. The period analysis of
PC2 in whole time interval (1974 - 2006) revealed the orbital
period $P_{\rm PC2} = (555.9 \pm 1.2)$ days. This orbital
period is also dominating in the period analysis of particular
stages (Q4, D, Q5 and E) except D, where the detected period is
$ \approx 480$ days. The period analysis of PC2 confirmed also
the presence of the shorter period $ \approx 340 - 370$ days in
all particular stages, but with lower significance than in PC1.
In the light curve of PC2 (see Fig. \ref{PCA}) faint
suggestions of outbursts in stages D and E are also visible.
The period analysis of PC3 did not give the presence of any
significant period.

\begin{table}
\centering \caption{The results of period analysis of principal
component (PC1) for particular stages in 1974 - 2006.}
\label{varPrincCompPC1}
\begin{tabular}{ll}
\hline
ID & Significant periods [days] \\
\hline
Q4 & $563.9 \pm 11.2; 343.0 \pm 5.6$ \\
D  & $484.5 \pm 7.2; 367.9 \pm 4.1$ \\
Q5 & $575.6 \pm 11.5; 348.1 \pm 3.8$ \\
E  & $367.9 \pm 1.1; 526.1 \pm 3.6 $ \\
\hline
\end{tabular}
\end{table}

The activity of the system is manifested in differential $
R_{\rm i}$ photometry, too. From all observational nights, the
data with magnitudes in all four filters were selected. For the
98 observations, the variation analysis as well as the period
analysis of principal components were performed. The variation
analysis showed that during the active stages the light curve
in $\Delta R_{i}$ band correlated well with LCs in all another
bands. For correlation coefficients, the following values were
obtained: $C_{\rm RU} = 0.95$, $C_{\rm RB} = 0.97$, $C_{\rm RV}
= 0.97$. It means that the outbursts are dominated in the red
light curves, too. The period analysis of the principal
component PC1 disclosed the significant presence of the
"postulated" pulsation period $P_{\rm pul} = (352.4 \pm 6.3)$
days. The next significant period $P_{\rm orb} = (544.2 \pm
17.4)$ days is close to the value determined for the orbital
period. In the light curves PC2 and PC3, only unreal periods
around 700 days were found. The last component PC4 did not
reveal any significant periods.

\section[]{Period analysis of spectroscopic data}
For the period analysis of the radial velocities, we used the
method of Fourier harmonic analysis \citep{andronov_1994},
which fits the first harmonic term of a trigonometric
polynomial approximation and Variable Stars Calculator (Breus
2003), which utilised Lafler-Kinman-Kholopov method. For
verification of the results, the method of Fourier analysis
\citep{ghedini_1982} was used.

\subsection[]{Period analysis of absorption lines}
\begin{figure}
\begin{center}
\includegraphics[width=0.45\textwidth]{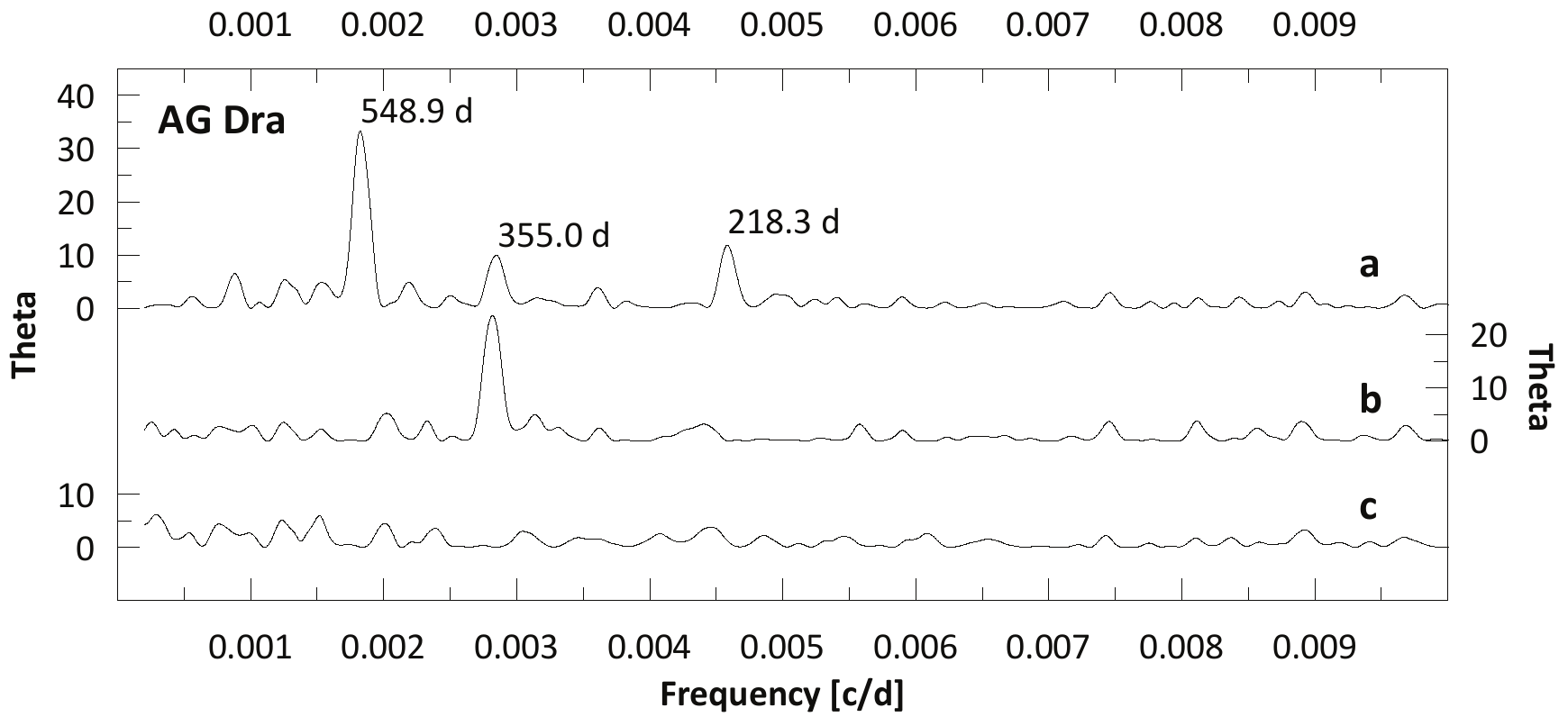}
\caption{Power spectra of AG Dra taken from combined radial
velocities based on absorption line measurements: original data
(a), orbital response removed data (b) as well as orbital and
probable pulsation response removed data (c).}
\label{ag_perRV}
\end{center}
\end{figure}

In our previous papers \citep{galis_1999, friedjung_2003} we
performed detailed period analysis of radial velocities based
on absorption lines measurements. We would like to note that
these measurements have very high accuracy with typical errors
(0.4 - 0.8 km/s). The data (135 radial velocities) cover the
time interval JD 2~446~578.5 - 2~451~676.9 (5098 days). We
reanalysed all these data to confirm the presence of periods
longer than 1000 days. We detected only two significant periods
$550.4 \pm 1.4$ and $355.0 \pm 1.6$ days related to the orbital
motion and cool component pulsations, respectively. The power
spectra taken from combined radial velocities based on
absorption line measurements are depicted in Fig.
\ref{ag_perRV}. We can confirm that the data did not contain
variability with longer periods. Moreover, the statistical
analysis suggests that data with those errors did not contain
other signal except orbital motion and postulated pulsation
ones.

\subsection[]{Period analysis of emission lines}
We have obtained equivalent widths (EW) and absolute fluxes of
spectral emission lines $\rm H_{\rm \alpha}$, $\rm H_{\rm
\beta}$, He~I~6678 \AA, He~II~4686 \AA and Raman scattered O VI
6825 \AA~as well as radial velocities of these lines (except
the O VI 6825 \AA~line). The variability of EW of some spectral
lines of AG~Dra is well known already for many years, and many
authors have made a great effort to assign this variability to
present physical mechanisms in the system \citep{kaler_1987}.
Up to now, this variability has not been explained without
uncertainties. The variation of EW is related to the increase
or decrease of the amount of observed emitted/absorbed
particles. This variability would be related to the orbital
motion in dependence of orientation of the system towards the
line of sight in a given moment (i.e. orbital phase). For more
details see \citet{leedjarv_2004}.

\begin{figure*}
\begin{center}
\includegraphics[width=0.95\textwidth]{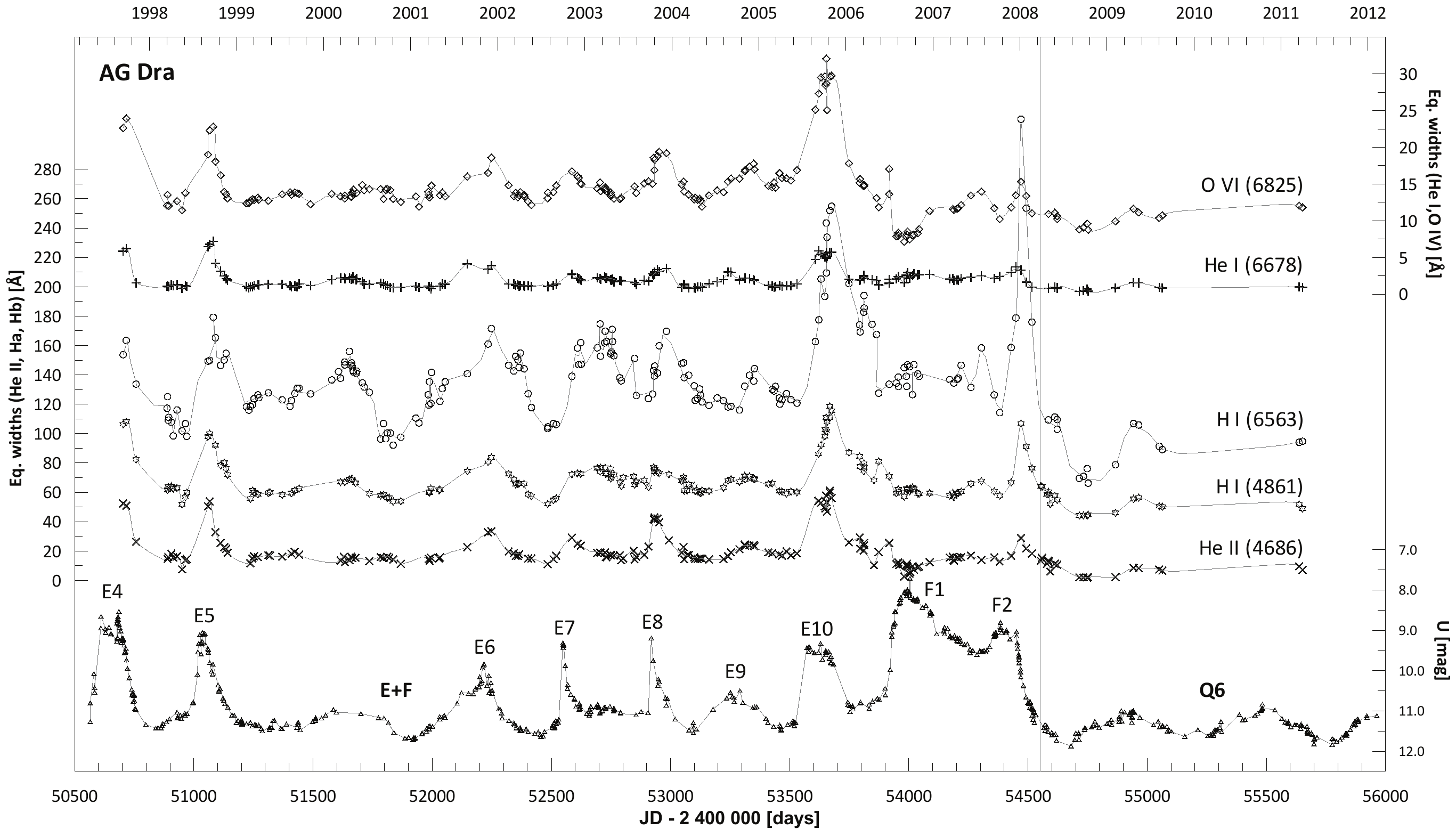}
\caption{The curves of EW for particular spectral lines. The
scales on the left and right axis are valid for EWs of $\rm He
\, I \ (6678 \, \rm \AA)$ and $\rm He \, II \ (4686 \,  \rm
\AA)$, respectively. The thin line shows a spline fit to the
data points.} \label{ag_ew}
\end{center}
\end{figure*}

The results of the period analysis of this data can be
summarized as follows: The most significant periods detected in
radial velocities, fluxes and EWs for all emission lines are
close to the orbital period (511 - 568 days) and to the time
interval between individual outbursts (366 - 383 days). The
period related to the pulsation of the red giant (350 - 357
days) was marginally detected. The period analysis did not show
the presence of the longer significant periods around 1160 days
like as the analysis of radial velocities based on the
absorption lines measurements (see previous section). There are
only a few detections of an insignificant period around 1100
days what is the double of the orbital period. The curves of EW
for particular spectral lines are depicted in Fig. \ref{ag_ew}.

\section{Discussion and conclusions}

The main goal of this paper was the complex and detailed period
analysis of photometric and spectroscopic data of AG~Dra. The
photometry covers the time interval of 124 years, while the
last 39 years of photometric observations are based on
systematic photoelectric and CCD monitoring. Spectroscopic data
were obtained from absorption and emission lines measurements.
The results of period analysis of all this data are two real
periods present in this symbiotic system: 550 and 350 days
related to the orbital motion and postulated pulsation of the
cool component, respectively. The orbital period is mainly
manifested during the quiescent stages at the shorter
wavelength ($U$ band), the pulsation period is present during
the quiescent as well as active stages at longer wavelengths
($B$ and $V$ bands). The period analysis of active stages
confirmed the presence of around 365 days period what is median
of the time intervals between outbursts. It is worth to note
that these time intervals vary from 300 to 400 days without
apparent long-term trend. The period analysis of the active
stages also revealed longer periods (e. g. 1330, 1580, 2350,
5500 days). Our detailed analysis shows that most of these
periods are more likely related to the complex morphology of
the light curves during the active stages than to the real
variability present in this symbiotic system.

Recently, the properties of the light curve of AG Dra covering
a period of 120 years were studied by \citet{formiggini_2012}.
Their research was mainly based on photographic observations
(Robinson 1969) and visual estimates from the AAVSO database.
It is worth noting that analysis of such data could give unreal
periods (e. g. 400, 440 days periods during Q2 and Q3 stages in
our analysis). As complementary observations,
\citet{formiggini_2012} used brightness measurements of AG~Dra
in the filter $U$ from literature. In our opinion, their
proposed model is unrealistic and non-physical in some aspects.
The result of the LC period analysis of these authors is the
detection of the period 373.5 days, that is a mean time
interval between the individual outbursts in the active stages.
Authors interpreted this period as the synodic rotational
period of the cool giant with respect to the white dwarf (for a
given point on the giant surface is a white dwarf in the upper
culmination every 373.5 days). To secure such synodic
rotational period in the binary with the orbital period around
550 days, the giant should rotate retrogradely with a period of
1160 days. The detection of this period is also reported by
these authors. We could not confirm the presence of this period
in photometric as well as in spectroscopic data. Moreover, such
value of the rotational period of the giant is not typical in
symbiotic systems (e. g. Table 2, in \citet{formiggini_2012})
and explanation of the retrograde rotation of a component in
such an open system from the evolutionary point of view is
unclear.

\citet{formiggini_2012} suggest that the cool giant of AG Dra
has a very strong magnetic field whose axis is substantially
(around 90 degrees) inclined relative to the rotational axis.
Moreover, they suggest that the outbursts of AG~Dra are the
result of intensive outflow of matter from the giant towards
the white dwarf. When a region of the magnetic poles of the
giant gets to the tidal bulge (strong tidal deformation of the
shape of a giant in the direction toward the white dwarf), the
balance is disrupted and hydrogen rich matter is thrown into
the Roche lobe of the white dwarf. This will release the large
amounts of gravitational energy which becomes apparent as
outburst observed in the optical. Since the amount of released
matter depends on the intensity of the magnetic field, as well
as hydro and thermo-dynamic properties of matter in the tidal
bulge at a given time, each outburst can vary greatly.
According to the opinion of specialists studying the stellar
magnetic activity across the HR diagram, such very strong
magnetic fields of cool giant are not known
\citep{korhonen_2013}. We do not understand the process of
balance breaking by a tidal bulge, in view of the fact that the
whole surface of the tidally deformed giant in a binary lies on
the same equipotential surface. The authors explain the
alternating active and quiescent stages of AG~Dra by the
mechanism similar to the solar dynamo (responsible for the
solar cycle activity), which takes place in the outer layers of
the extensive giant atmosphere. Our analysis of all 124 years
brightness history has shown that active stages have
reccurrency 12 - 16 years which is in good agreement with the
idea about solar-like activity.

The period analysis of photoelectric and CCD observations
obtained during the active stages D and E + F supported the
results of our statistical analysis of the data: the light
curve in the $U$, $B$ and $V$ bands are strongly
cross-correlated during the activity of AG~Dra. The obtained
periods in individual bands have the same values within their
errors. Global morphology of the active stages is well
described by two longer periods (about 1500 and 5400 days). The
obtained power spectra contained also a period of about 1 100
days, but the significance of the period decreased considerably
after removal of these two longer periods from the light curves
of AG~Dra (Fig. \ref{AG_perUBV}). From this fact we suggested
that the 1100 days period more likely describes only the
morphology of the active stages of AG~Dra than a real
variability presented in this system. The obtained power
spectra after removal of the longer periods (about 1500 and
5400 days) showed the presence of a significant period around
375 days (Fig. \ref{AG_perUBV}), which is related to the
outburst time distribution. As can be seen, the maximum of
significance of this period has two peaks (second peak around
360 days), suggesting a possible long-term change of its value.
This assumption was confirmed by detailed period analysis of
individual parts of the active stage E + F. While at the
beginning of E+F stage the power spectrum contains only a
period of about 370 days; at the end of this active stage, the
power spectrum contains only a period of 360 days. The
detection of the period of about 559 days (close to the value
of the orbital one), which was significant during the stages of
activity in $B$ and $V$ bands is also interesting.

The situation was different during the quiescent stages (Q4 -
Q6) and the period analysis confirmed that the light curve in
the $U$, $B$ and $V$ bands did not cross-correlate. The period
analysis clearly confirmed the presence only of the orbital
period (551.8 days) in the $U$ band as well as the orbital
(551.7 days) and pulsation (351.2 days) periods in $B$ band
(Fig. \ref{AG_perUBV}). The power spectrum in $V$ band (Fig.
\ref{AG_perUBV}) is more complex: although the orbital (547.2
bottom) and pulsation (350.6 bottom) periods were detected, the
resulting power spectrum also contained other significant
periods (334.1, 223.0 and 136.8 days). Though this may be only
an artefact caused by the fact that amplitude of the light
variations in the $V$ band during quiescent stages of AG~Dra is
comparable to the observation errors in this band.

The interpretation of the time intervals between outbursts with
a median of 365 days is more complicated. The emission spectral
lines are created in the part of the symbiotic nebula,
originating from the stellar wind of the red giant and ionized
by radiation from the white dwarf. The stellar wind should be
modulated by the giant pulsations. This modulation produces a
Doppler shift of the spectral lines, that is manifested in
radial velocities. Simultaneously also the number of
emitted/absorbed particles, changes what is possible to detect
in EW variability. \citet{skopal_2009} proposed another model,
where the hydrogen lines are created in the wind of the hot
component. In this case, the main role play is the mutual
interactions of the wind on both components, but in this model,
it is impossible to explain all observed effects.

\citet{leibowitz_2006, leibowitz_2008, leibowitz_2011,
leibowitz_2013} and \citet{formiggini_2006, formiggini_2012}
have found similar patterns in long-term light curves of six
symbiotic stars, including AG~Dra. The common feature of all
these stars is the start of active periods at about 4650--7550
day intervals (12--20 years). This implies a common physical
mechanism, for example, solar-like magnetic cycles, might be
responsible for the activity of these symbiotic stars. However,
there is no direct evidence of the presence of a reasonably
strong magnetic field in the red giants in these systems.
Complex morphology of the light curves may introduce artefacts
in their period analysis. We suggest that a careful case by
case analysis of the light curves together with spectroscopic
data and when possible, observations in the wide wavelength
range from X-rays to radio, would give first clues to
understanding the physical mechanism of the outburst behaviour
of symbiotic stars. One should not pay too much attention to
the possible artificial periods obtained from the analysis of
the complicated light curves. The outburst phenomena are not
necessarily strictly periodic as we have seen in the case of
AG~Dra. Let us be reminded that all the active periods of
AG~Dra have been different from each other (Figs. \ref{ag_b}
and \ref{UBV}).

One of the promising explanations of at least some individual
outbursts of AG~Dra might be the combination nova model
proposed for Z~And by \citet{sokoloski_2006}. In this model,
when accretion rate onto the white dwarf exceeds some critical
value, thermonuclear reactions are ignited and luminosity of
the hot component increases significantly. One of the next task
would be to study whether hot and/or cool outbursts of AG~Dra
will fit into such a picture. We will present a detailed
analysis of the spectroscopic data of AG~Dra in our forthcoming
paper.

\section*{Acknowledgments}We would like to devote this paper to the memory of
our friend Michael Friedjung (1940--2011), long-term
collaborator on AG~Dra and other symbiotic stars. Michael
suddenly passed away but we are sure that his wonderful spirit
is now together with us.

This study was supported by the Slovak Academy of Sciences VEGA
Grant No. 2/0038/13 and by the Estonian Ministry of Education
and Research target financed research topic SF0060030s08. This
article was also supported by the realisation of the Project
ITMS No. 26220120029, based on the supporting operational
research and development program financed from European
Regional Development Fund.

\label{lastpage}

\end{document}